\documentclass[pdflatex,sn-mathphys-num]{sn-jnl}

\usepackage{graphicx}%
\usepackage{multirow}%
\usepackage{amsmath,amssymb,amsfonts}%
\usepackage{amsthm}%
\usepackage{mathrsfs}%
\usepackage[title]{appendix}%
\usepackage{xcolor}%
\usepackage{textcomp}%
\usepackage{manyfoot}%
\usepackage{booktabs}%
\usepackage{algorithm}%
\usepackage{algorithmicx}%
\usepackage{algpseudocode}%
\usepackage{listings}%

\usepackage{amsthm}
\usepackage{lmodern}

\begin{document}

\title[Article Title]{Qubit Optimized Quantum Implementation of SLIM}


\author*[1,2,3]{\fnm{H. O. } \sur{Cildiroglu}}\email{hoc@physics.bu.edu}

\author[3]{\fnm{O.} \sur{Yayla}}\email{oguz@metu.edu.tr}

\affil[1]{\orgdiv{Physics Department}, \orgname{Boston University}, \orgaddress{\street{Commonwealth Ave}, \city{Boston}, \postcode{02100}, \state{MA}, \country{USA}}}

\affil[2]{\orgdiv{Department of Physics Engineering}, \orgname{Ankara University}, \orgaddress{\street{Dögol st}, \postcode{06100}, \state{Ankara}, \country{Turkiye}}}

\affil[3]{\orgdiv{Institute of Applied Mathematics}, \orgname{Middle East Technical University}, \orgaddress{\street{Universiteler}, \city{Ankara}, \postcode{06800}, \state{Ankara}, \country{Turkiye}}}


\abstract{The advent of quantum computing has profound implications for current technologies, offering advancements in optimization while posing significant threats to cryptographic algorithms. Public-key cryptosystems relying on prime factorization or discrete logarithms are particularly vulnerable, whereas block ciphers (BCs) remain secure through increased key lengths. In this study, we introduce a novel quantum implementation of SLIM, a lightweight block cipher optimized for 32-bit plaintext and an 80-bit key, based on a Feistel structure. This implementation distinguishes itself from other BC quantum implementations in its class (64–128-bit) by utilizing a minimal number of qubits while maintaining robust cryptographic strength and efficiency. By employing an innovative design that minimizes qubit usage, this work highlights SLIM’s potential as a resource-efficient and secure candidate for quantum-resistant encryption protocols.
}




\maketitle 
\section{Introduction}\label{sec1}

Classical cryptography methods primarily hinge on exploiting computational inefficiencies within mathematical problems such as prime factorization and discrete logarithms to uphold the security of encrypted data. The potential of quantum computers to efficiently solve these problems, which pose significant challenges for classical computers, using quantum algorithms fundamentally threatens the security of existing encryption protocols \cite{shor1997, grover1996, yamamura2000, simon1994, dowling2003, polkovnikov2011, harrow2017}. Therefore, there is a burgeoning interest in crafting algorithms resilient to quantum computing methods, ensuring secure communication and data protection \cite{daemen2001, li2022, huang2022, almazrooie2018, langenberg2020a, luo2022, langenberg2020b}. In this context, it is crucial to construct quantum algorithms for classical encryption methods, strengthening them against potential threats and thoroughly assessing their resilience to quantum attacks \cite{selinger2013quantum, dasu2019, anand2020, saravanan2021, jing2023, paramasivam2023, luo2024quantum, chun2023dorcis, lee2022}.

Block ciphers (BC) are still considered to be  quantum-resilient algorithms employed for encrypting data blocks or fixed-length bit groups. Most BCs are designed to encrypt data in predefined blocks of either 64 or 128 bits in size \cite{Feistel1973, DES1999, AES2001, jaques2020, schlieper2020, pal2022, chauhan2020, bogdanov2007, hong2006, knudsen2010, guo2011, beaulieu2015, zhang2014, cheng2008, wu2011, patil2017}. The Feistel structure is a fundamental design element of symmetric BC algorithms, enabling the partitioning of blocks, the use of subkeys, round-based operations, and an iterative structure to perform both encryption and decryption processes with the same key \cite{Feistel1973, DES1999}. Designed to enhance encryption security, it provides a flexible framework that can be utilized in many encryption algorithms. SLIM, a novel variant of the Feistel-based BCs with 32-bit block size, is designed for lightweight applications from contemporary RFID technologies to the Internet of Things (IoTs), striking a balance between security and efficiency \cite{aboushosha2020}. In the construction of SLIM, four 4x4 S-boxes are utilized to function as a non-linear component of the cipher, executing a non-linear operation on a 16-bit word. In this regard, SLIM is proposed to be secure, despite its low bit usage compared to other BCs.

In this work, we present a quantum implementation of SLIM with the aim of developing encryption methods that are robust against emerging quantum technologies. Achieving an optimized quantum cost, our implementation employs the smallest number of qubits compared to other BC quantum implementations in existing literature. To achieve this, we conduct a comprehensive analysis of the SLIM algorithm in the context of quantum circuits, focusing on its Key Addition-Substitution-Permutation (KSP) layers. By strategically reversing the KSP structure of SLIM and avoiding the use of ancilla qubits, we successfully construct for the first time the algorithm with a total of 112 qubits and a quantum cost of 27220. This approach lays the foundation for a practical implementation in current technology, providing an opportunity to test and enhance its resilience in contrast to quantum attacks. 

This paper is organized as follows. In Section II, we present the KSP structure of the SLIM, along with the notation designed for quantum circuits. In Section III, we perform the quantum implementation of SLIM using quantum gates. The fourth and final section is dedicated to calculating the quantum costs associated with the implementation of SLIM and discussing it in comparison with other BC implementations.

\section{The structure of SLIM}\label{sec2}

The development of quantum technologies poses a significant threat to the security of existing cryptographic algorithms, encouraging the exploration of quantum-resilient (or post-quantum) cryptography. In this regard, BC stands out with its simple but resilient design. However, some BC algorithms may be solved by quantum computers due weakness in their design, eg Grover search attack on their differential characteristics  \cite{yadav2023}. 
On the other hand, SLIM, a recently introduced variant of BC, promises to be a quantum-resistant algorithm with its lightweight design and a well-balanced Feistel structure in terms of security and efficiency \cite{aboushosha2020}. 

\begin{figure}[h]
\centering
\includegraphics[width=0.9\textwidth]{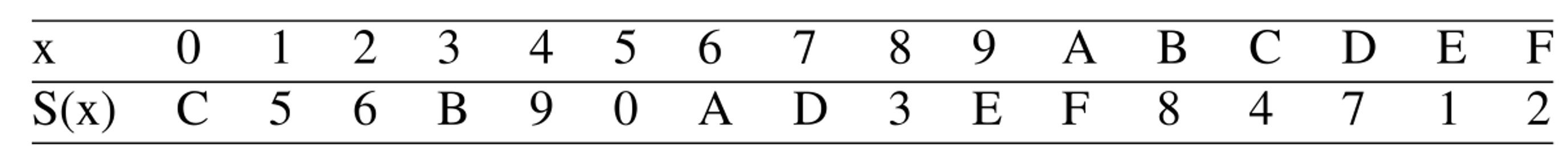}
\caption{\centering S-box of SLIM} 
\label{fig1}
\end{figure}

SLIM is a symmetric encryption algorithm which has Feistel structure that uses the same key for both encryption and decryption. The only difference between these processes is the use of decryption sub-keys in reverse order. SLIM integrates both confusion and diffusion principles. Confusion is efficiently handled through a compact 4-bit S-box, using the same structure of PRESENT \cite{bogdanov2007}. Simplicity is a crucial aspect of SLIM, evident in the compact size of the S-box (Fig.~\ref{fig1}) and the use of inherently straightforward operations. Operating with an 80-bit key, SLIM is designed to encrypt and decrypt 32-bit plaintext and ciphertext blocks.

\begin{figure}[h]
\centering
\includegraphics[width=0.9\textwidth]{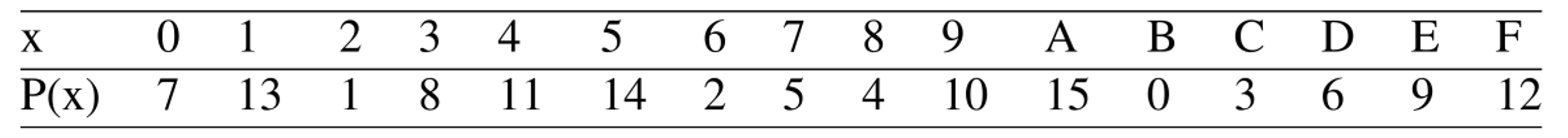}
\caption{\centering P-box of SLIM} 
\label{fig2}
\end{figure}

\noindent The rearrangement phase of SLIM utilizes a crucial permutation layer, generating a 16-bit output from 16-bit inputs through a meticulously selected rule specified in Fig.~\ref{fig2}. This choice as mentioned in its proposal \cite{aboushosha2020} arises from a need for securing against both linear and differential cryptanalysis. In linear cryptanalysis, the permutation rule is designed to resist patterns, creating substantial confusion to hinder adversaries from identifying patterns within the cipher. Simultaneously, the rule takes into account differential cryptanalysis, disrupting systematic relationships between input and output differentials to fortify the cipher's resilience against such attacks. 
Hence, these techniques play a role in thwarting various cryptanalytic approaches, illustrating the algorithm's effectiveness and suitability for secure information transmission.

\begin{figure}[h]
\centering
\includegraphics[width=1\textwidth]{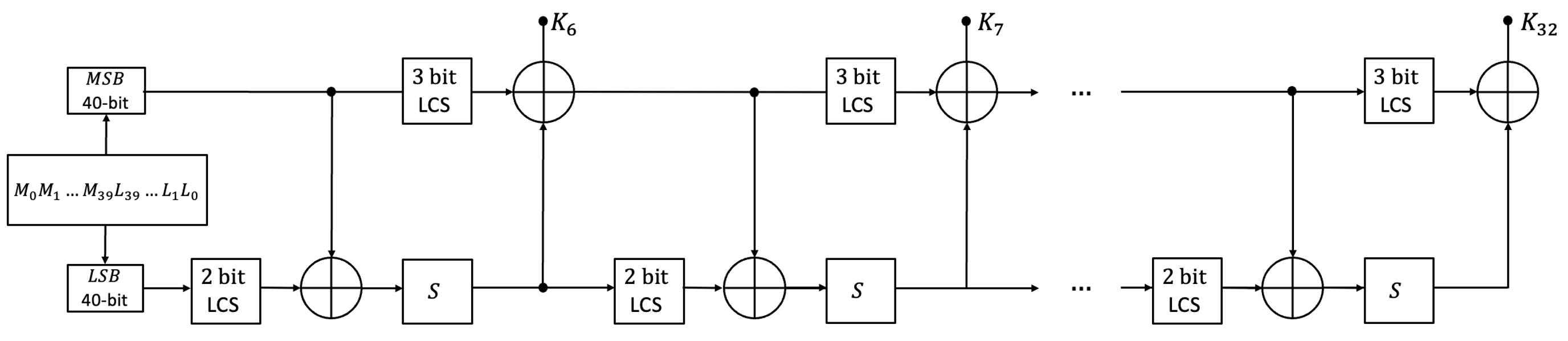}
\caption{\centering Key Generation of SLIM $(i>5)$.} 
\label{fig3}
\end{figure}

\noindent The 32 sub-keys (16-bit each), required for 32 rounds, are generated from the 80-bit encryption key in SLIM. NIST guidelines recommend a key length of at least 80 to resist exhaustive search attacks \cite{nist2016}. The initial five sub-keys $\{K_1,\dots,K_5\}$ are directly derived from the original key ($K = M_0 M_1 \dots M_{39} L_{39} \dots L_1 L_0$). Specifically, $K_1$ corresponds to the first least significant 16 bits, $K_2$ to the subsequent 16 bits, and so forth (up to the fifth round). The 80-bit key is then divided by a splitter, yielding two 40-bit values as the most and least significant bits (MSB and LSB). Each half is subsequently processed individually (See Fig.~\ref{fig3}). In each round, the LSB undergoes a left cyclical shift (LCS) of two bits, followed by an XOR operation with the MSB. The output of this XOR process is then passed to a substitution layer. The round sub-key is created by manipulating the output of the S-boxes and the rotated MSB, achieved through an LCS of 3 bits using the XOR technique. 

\begin{figure}[h]
\centering
\includegraphics[width=1\textwidth]{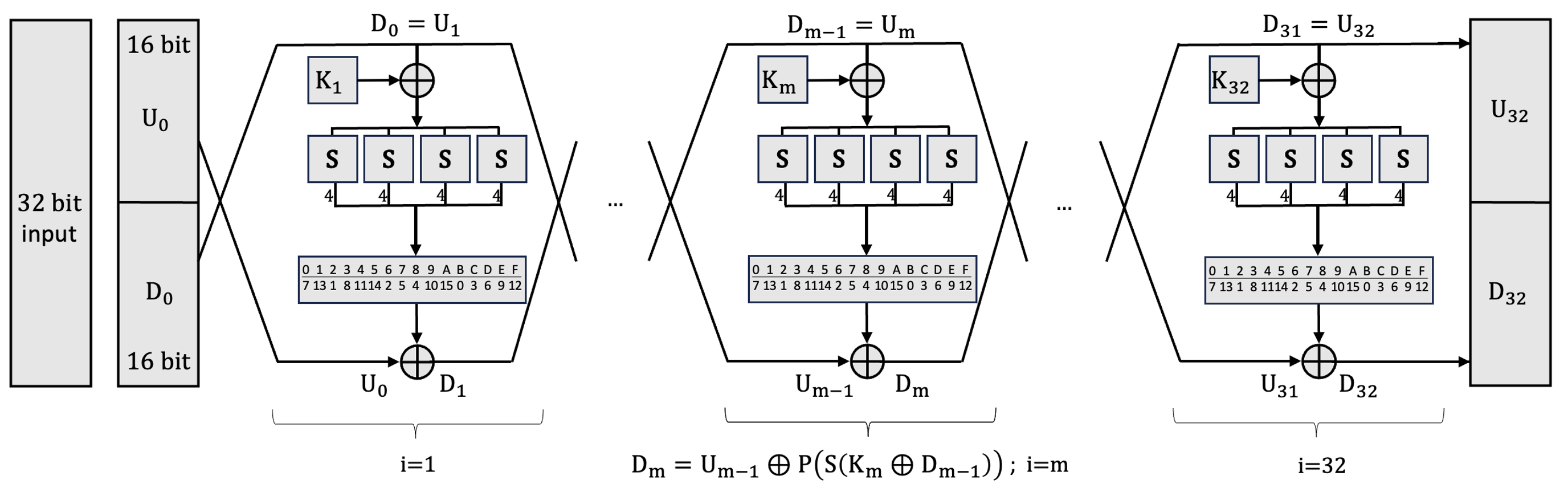}
\caption{\centering P-box of SLIM} 
\label{fig4}
\end{figure}

\noindent In the SLIM's encryption algorithm, the input splits into up and down parts, which go through 32 rounds of processing along with the generated sub-keys. 
The interior structure of SLIM is given in Fig.~\ref{fig4}. According to this, the 32-bit input is split into two equal sixteen-bit halves as $U_i$ and $D_i$, where $U_i = D_{i-1}$ and $D_i = U_{i-1} \oplus P(S(K_i \oplus D_{i-1}))$. 

\section{Quantum Implementation of SLIM}\label{sec3}

In quantum mechanics, the state of a system is described by wave functions ($|\psi\rangle$), defined in the Hilbert space ($\mathcal{H}$)-a vectorial complex inner product space. These wave functions represent the probability amplitude of a particle that exhibits a specific property (e.g. position or spin) upon measurement. Thus, a quantum state is a complex linear superposition of possible substates within this space. Single-particle quantum systems with two subparts, such as horizontal $|H\rangle$ and vertical $|V\rangle$ polarization, spin-up $|\uparrow\rangle$ and spin-down $|\downarrow\rangle$, or more broadly $|1\rangle$ and $|0\rangle$, are referred to as quantum bits (or simply qubits). Qubits are the quantum counterparts of classical bits, the fundamental units of classical computation. The general state of a single qubit is expressed as:

\begin{equation}
    |\psi\rangle = a|1\rangle + b|0\rangle; \quad |\psi\rangle \in \mathcal{H}_2; \quad a, b \in \mathbb{C}; \quad a^2 + b^2 = 1.
\end{equation}

\noindent   Composite systems are formed by combining two or more discrete and separately prepared qubits ($|\psi^{(i)}\rangle \in \mathcal{H}^i ; i = 1,\dots,n$), which reside in the tensor product space $\mathcal{H}^{\otimes n}$, can be given as:

\begin{equation}
    |\Psi\rangle = |\psi^{(1)}\rangle \otimes \dots \otimes |\psi^{(n)}\rangle \quad \text{(or simply } |\psi^{(1)} \dots \psi^{(n)}\rangle\text{)}, 
\end{equation}

\noindent where $|\Psi\rangle \in \mathcal{H} = \mathcal{H}^{(1)} \otimes \dots \otimes \mathcal{H}^{(n)}$. Besides, due to the causality of the theory, the evolution of the system can be controlled by unitary operators such as $|\Psi' \rangle = \hat{U} |\Psi\rangle$. If the system undergoes a stepwise evolution, its state is transformed sequentially by a series of unitary operators. The final output state of the system is $|\Psi' \rangle = \hat{U}_n \dots \hat{U}_2 \hat{U}_1 |\Psi\rangle$ or, equivalently,

\begin{figure}[h]
    \centering
    \includegraphics[width=0.4\linewidth]{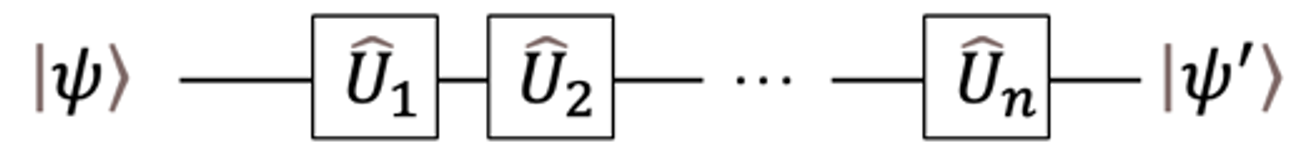}
\end{figure}

\noindent On the other hand, the transformed state of the $n$-qubit system is $|\Psi' \rangle = \left[\hat{U}^{(1)} \otimes \dots \otimes \hat{U}^{(n)} \right] \left[|\psi^{(1)}\rangle \otimes \dots \otimes |\psi^{(n)}\rangle \right]$ or equivalently,

\begin{figure}[h]
    \centering
    \includegraphics[width=0.4\linewidth]{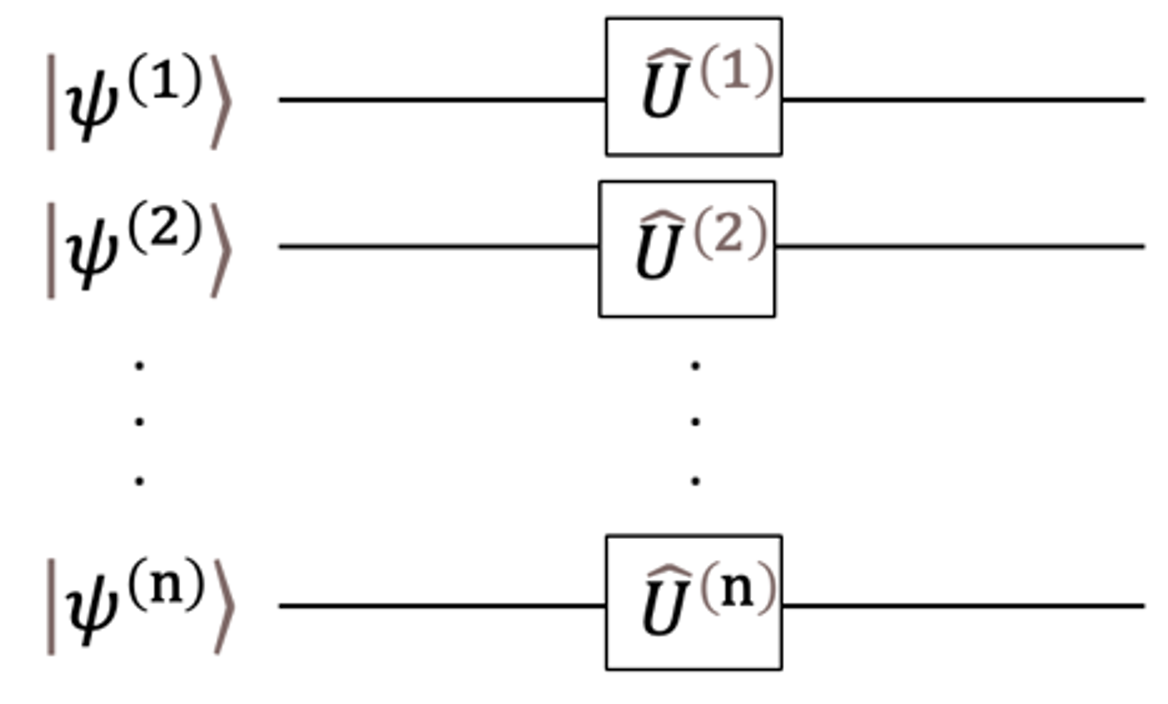}
\end{figure}

\begin{figure}[h]
\centering
\includegraphics[width=1\textwidth]{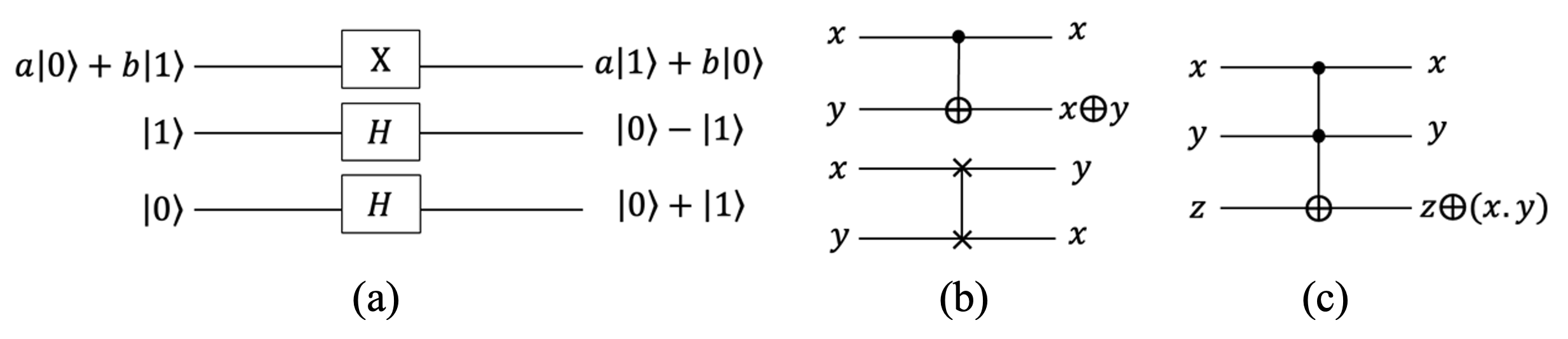}
\caption{\centering Quantum gates: (a) X: NOT, H: Hadamard; (b-top) CNOT: Controlled-NOT, (b-bottom) SWAP; (c) CCNOT. Here $a, b \in \mathbb{C}$, $x, y, z \in \{0,1\}$, and $\oplus$ is the bitwise XOR.
} 
\label{fig5}
\end{figure}

\noindent Here, the evolution operator $\hat{U}^{(i)}$ acts on the ket $|\psi^{(i)}\rangle$. These representations are suitable for constructing quantum implementations of the S, P, and K layers for the encryption algorithm given in the previous section. To achieve this, some quantum gates such as the X (or simply NOT), Hadamard ($H$) gate, CNOT, CCNOT (or Toffoli), and SWAP gates need to be introduced (See Fig.~\ref{fig5}).

To implement SLIM as a quantum circuit, the 32-qubit input is first divided into two equal parts: the upper half $(U)$ and the lower half $(D)$. The 16 qubits ($D_0$) from the D-box are combined with the first round key ($K_1$), derived by dividing the 80-bit key into five equal parts. Next, the S-box is executed four times in parallel, starting from the least significant qubits (LSQ). The resulting outputs are then passed through the P-box. For simplicity, this sequence of operations of key addition, substitution and permutation layers is referred to as KSP. Following the KSP process, the results are manipulated using CNOT gates with the 16 qubits ($U_0$) from the U-box, and the first round is completed by obtaining $D_1$.

At this stage, it is essential to ensure the feasibility of implementing the second and subsequent rounds of SLIM. When KSP is applied on $D_0$, it transforms into $U_1$. While this transition is straightforward from a classical perspective, it introduces challenges in quantum paradigm, as it necessitates duplicating the $D_0$ packet, which is a task that inherently requires additional qubits. This duplication involves adding and applying CNOT gates to a set of ancilla qubits, which are initialized to the $|0\rangle$ state, matching the size of the packet (16 qubits). Moreover, to complete all rounds of SLIM, new ancilla qubits must be introduced before each round, in addition to the initial set. In experimental setups where an increase in qubit count is negligible, such as idealized scenarios, this approach proves advantageous, as it significantly reduces the number of quantum gates required. However, given the limitations of current quantum computing technology, such a method is impractical due to the cost of qubit resources. 

Instead, we employ a more efficient and a novel strategy that avoids excessive qubit usage. By leveraging the Feistel structure of SLIM, we retrieve the original qubits for subsequent rounds through the reverse application of the KSP process (KSP$^{-1}$), which consists of P$^{-1}$, S$^{-1}$, and K$^{-1} =$ K, applied in reverse order from the lower to upper branches. This approach effectively prepares the system for the second round and allows $U_i$ to be reused as $D_{i-1}$ in each subsequent round. Now, let us realise the quantum implementations of the K, S and P layers and their inverses respectively.

\begin{figure}
\centering
\includegraphics[width=1\textwidth]{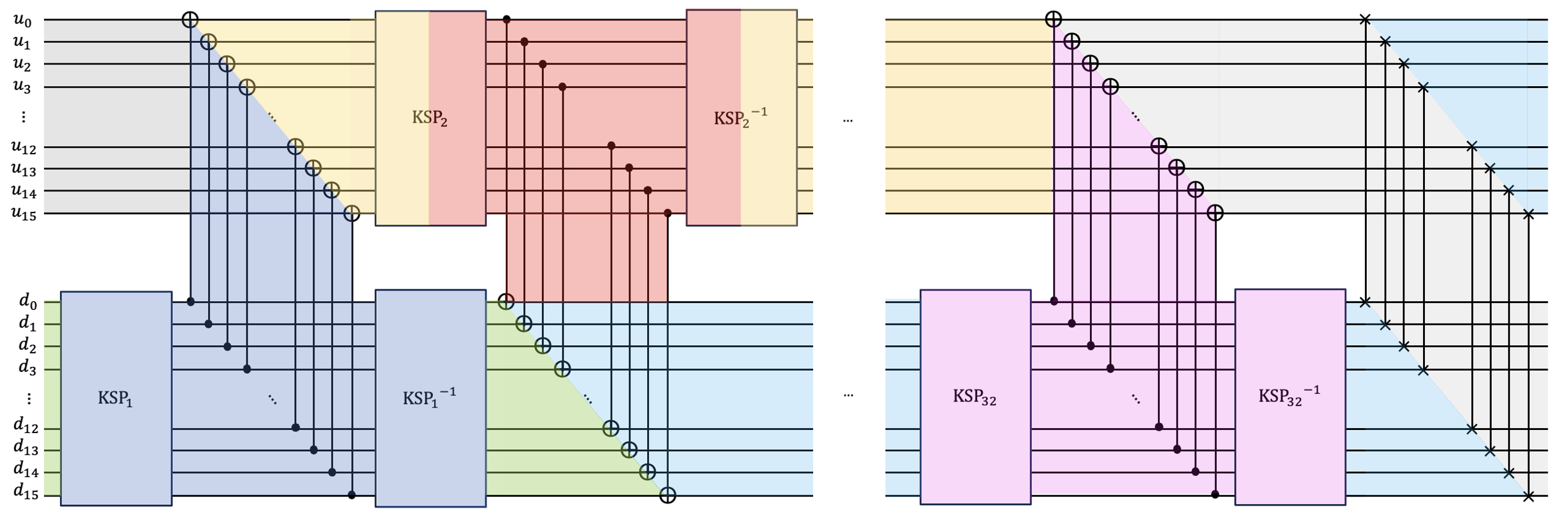}
\includegraphics[width=1\textwidth]{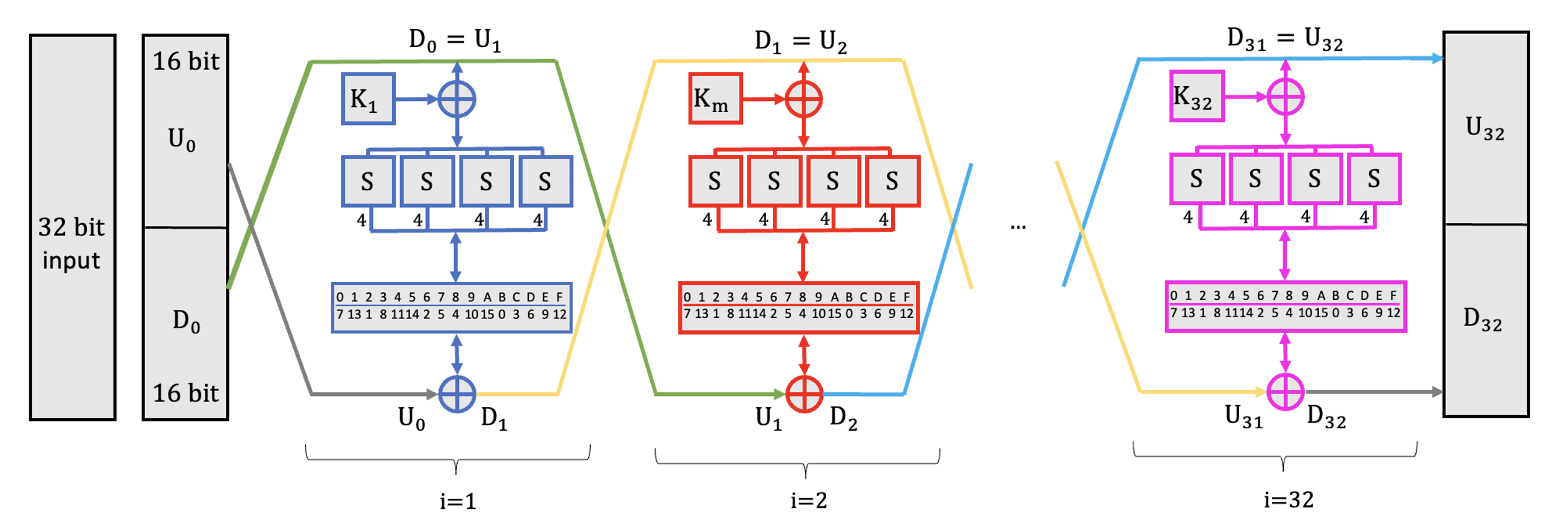}
\includegraphics[width=1\textwidth]{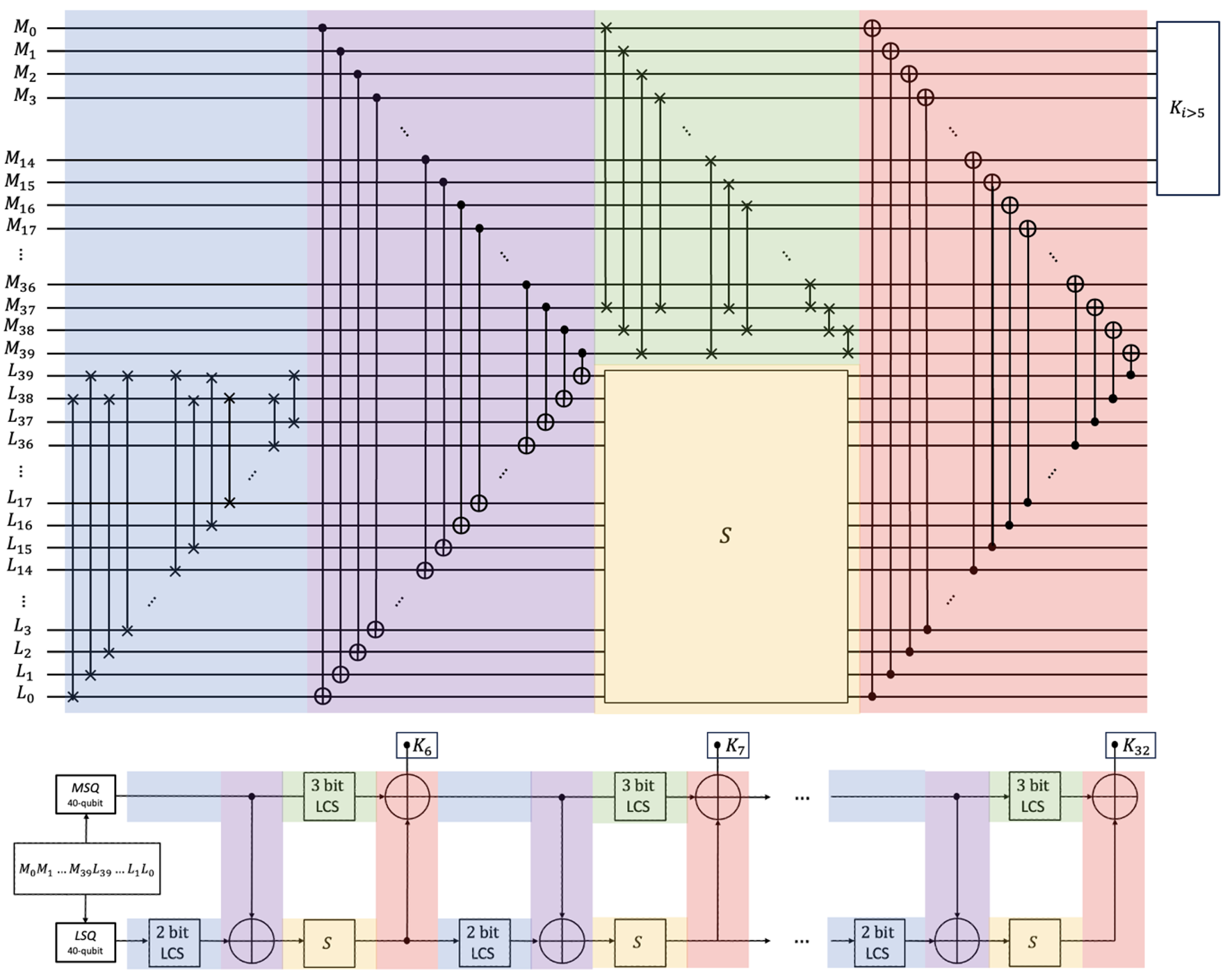}
\caption{\centering Quantum implementation of 32 rounds of SLIM and the key schedule for the rounds $i>5$} 
\label{fig6}
\end{figure}

\vspace{1em}

\noindent \textbf{Key-Addition Layer:} SLIM has 80 qubit keys and runs 32 rounds in total. In this context, for the first 5 rounds, the 80-qubit key is directly divided by five according to the LSQ and added as 16-qubit keys (See Fig.~\ref{fig6}). Then, the 80-qubit key goes through a divider to create two 40-qubit keys, labeled MSQ and LSQ. The resulting 40-qubit keys are then treated separately. Accordingly, in each round, the LSQ undergoes a two-qubit circular left shift operation (blue colored) and then the output produced is CNOTed with the MSQ (purple colored). The output of the CNOT process is forwarded to a substitution layer (yellow colored). The result of the S-boxes, which undergoes a three-qubit circular left-shift operation (green colored), is manipulated using the CNOT operations to generate the subkey (red colored). The first 16 qubits are then taken and used as the key. This cycle continues until the 32nd round.

\noindent The Feistel structure of SLIM inherently ensures that the inverse of the key addition layer, K$^{-1}$, is equivalent to the key addition layer itself, K. This property arises because the operations performed within the
K layer of CNOT, substitution, and circular shift are inherently reversible when applied in the same sequence. Consequently, for the decryption process, K$^{-1}$ is directly implemented as K, eliminating the need for additional computational overhead or resources to reverse the key schedule. This characteristic is a fundamental advantage of the Feistel structure, significantly simplifying the implementation of KSP$^{-1}$ during decryption.

\vspace{1em}

\noindent \textbf{S-box:} The S-box used in SLIM, as given in \eqref{eq:sbox}, is identical to the one designed for the PRESENT algorithm \cite{bogdanov2007}. Its quantum implementation has been optimized using the LIGHTER-R framework, a specialized tool for reversible circuit synthesis \cite{dasu2019}. LIGHTER-R is particularly well-suited for this task because it eliminates the need for ancillary qubits and minimizes garbage output while efficiently optimizing quantum gates. This tool provides an end-to-end solution for reversible S-box construction, offering significant advantages in terms of gate cost and resource efficiency. As such, the quantum circuits for both the S-box and its inverse, S$^{-1}$-box, are carefully designed in this work to align with these optimizations . To reconstruct the S-box as a quantum circuit with variable output, we use the quantum gates described in the preceding section (NOT, CNOT, and CCNOT gates). The Boolean functions that define the logical operations necessary for constructing the S-box transformations are as follows.

\begin{eqnarray} \label{eq:sbox}
x_0 &\rightarrow& x_3' x_2' x_0 \oplus x_3' x_1 x_0 \oplus x_3 x_2' x_0' \oplus x_3 x_1 x_0' \oplus x_3 x_2 x_1' x_0 \oplus x_3' x_2 x_1' x_0' \nonumber \\
x_1 &\rightarrow& x_3' x_2' x_1 \oplus x_3 x_2 x_0 \oplus x_3' x_1 x_0' \oplus x_3 x_2' x_1' \oplus x_3 x_2' x_0' \nonumber \\
x_2 &\rightarrow& x_3 x_2 x_1' \oplus x_2' x_1 x_0' \oplus x_3' x_2 x_1 x_0 \oplus x_3' x_2' x_1' \oplus x_2' x_1' x_0 \\
x_3 &\rightarrow& x_3 x_2' x_1 \oplus x_3 x_2' x_0 \oplus x_3' x_1' x_0' \oplus x_3' x_2 x_1 \oplus x_3' x_1 x_0 \nonumber
\end{eqnarray}

\begin{figure}[H]
\centering
\includegraphics[width=0.6\textwidth]{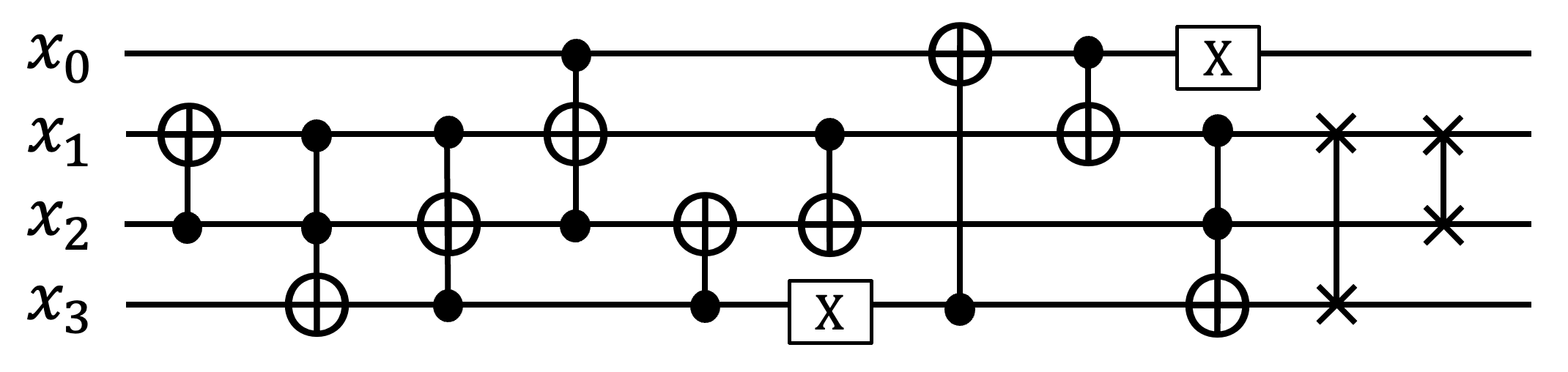}
\caption{\centering Quantum implementation of S-box} 
\label{Fig7}
\end{figure}

\noindent Here, the notation, e.g., $x_3' x_2' x_0$, indicates $[(\text{NOT } x_3) \text{ AND } (\text{NOT } x_2) \text{ AND } x_0]$. Using these functions and the LIGHTER-R framework (NCT-gc), the optimized quantum implementation of the SLIM S-box is constructed as shown in Fig.~\ref{Fig7}.

\noindent Similarly, the Boolean functions for the S$^{-1}$-box are as follows \cite{mohanapriya2023optimized}:

\begin{eqnarray}
x_0 &=& S_3' S_2' S_0' + S_2' S_1' S_0' + S_2 S_1' S_0 + S_3' S_2 S_0 + S_3 S_2 S_1 S_0' + S_3 S_2' S_1 S_0 \nonumber\\ 
x_1 &=& S_3' S_2' S_1' S_0 + S_3' S_1 S_0' + S_3 S_2 S_0 + S_3 S_1 S_0 + S_3 S_2' S_0' \nonumber\\
x_2 &=& S_3' S_2' S_1' + S_3' S_1' S_0' + S_3 S_1' S_0 + S_2' S_1 S_0' + S_3' S_2 S_1 S_0 \\ 
x_3 &=& S_3' S_2' S_0 + S_3' S_2' S_1 + S_2 S_1 S_0 + S_3 S_2 S_1 + S_3 S_2' S_1' S_0' + S_3' S_2 S_1' S_0' \nonumber
\end{eqnarray}

\begin{figure}[H]
\centering
\includegraphics[width=0.6\textwidth]{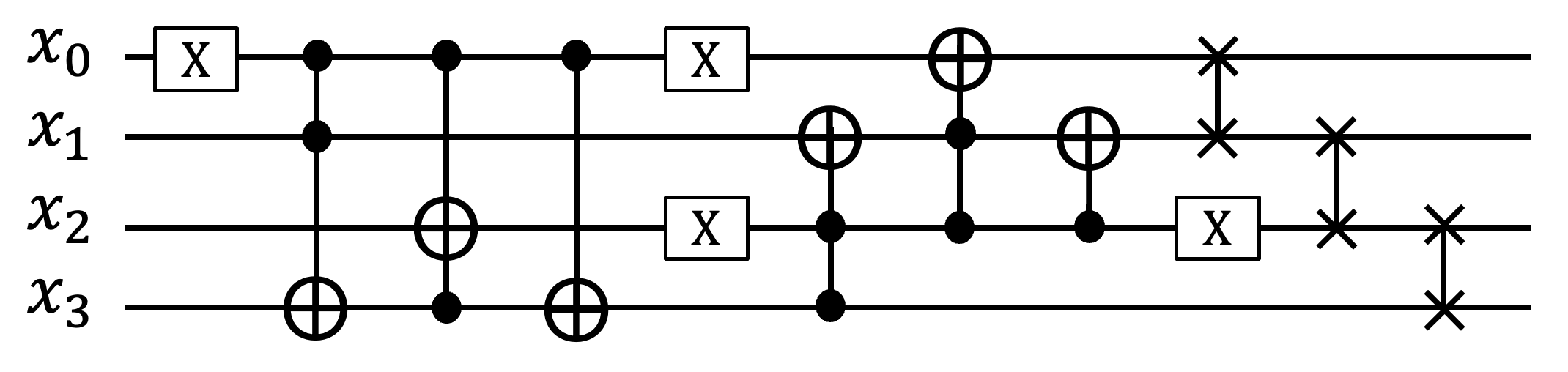}
\caption{\centering Quantum implementation of S$^{-1}$-box} 
\label{fig8}
\end{figure}

\noindent These functions, when implemented using the LIGHTER-R framework, allow for an efficient reversible construction of the S$^{-1}$-box. 
The resulting circuit (Fig.~\ref{fig8}) is optimized to balance quantum gate costs and qubit usage, ensuring compatibility with the overall KSP$^{-1}$ structure.

\noindent After the key addition layer processes the 16 bits of the input block $U_i$, the S-box operates in parallel for four 4-qubit segments, starting from the LSB. These transformed outputs are then routed through the P-layer, where the qubits are permuted according to a predefined mapping. 

\begin{figure}[H]
    \centering
    \begin{minipage}{0.45\textwidth}
        \centering
        \includegraphics[width=\textwidth]{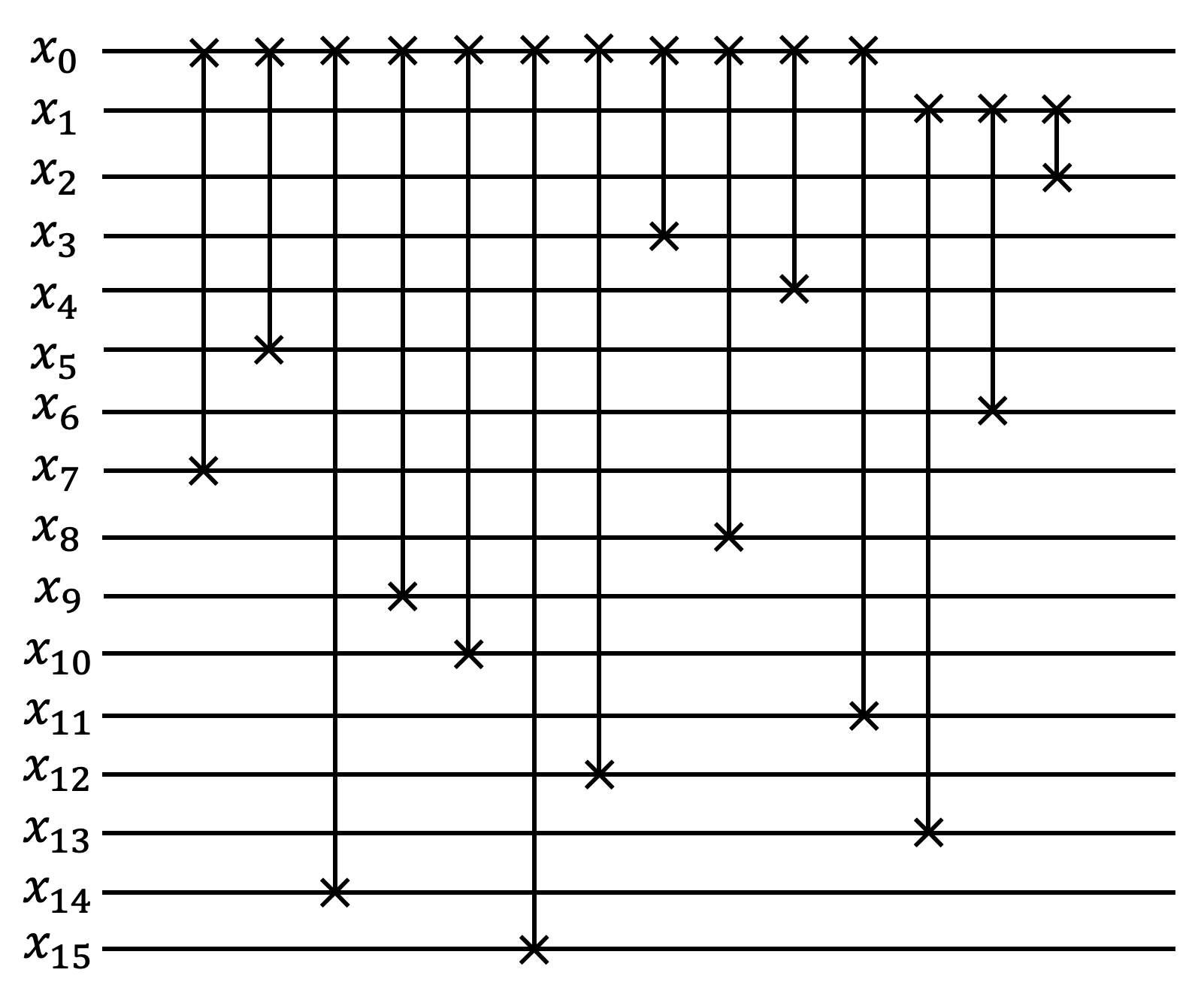}
    \end{minipage}
    \begin{minipage}{0.45\textwidth}
        \centering
        \includegraphics[width=\textwidth]{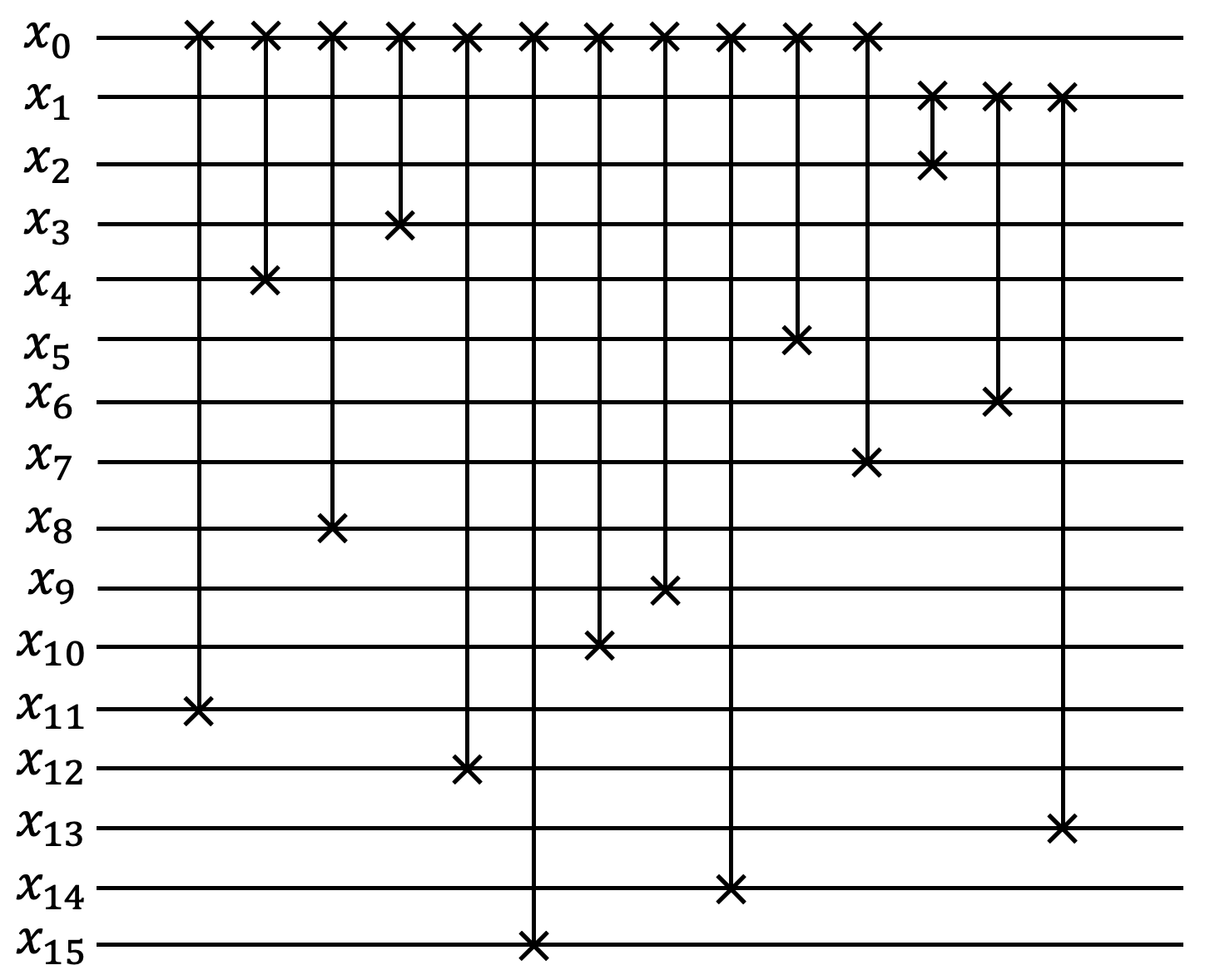}
    \end{minipage}
    \caption{\centering Quantum implementations of (a) P-box and (b) P$^{-1}$-box.}
    \label{fig9}
\end{figure}

\noindent\textbf{P-box:} The permutation layer given in Fig. ~\ref{fig2} can also be implemented in a quantum circuit. For this, it is sufficient to use the SWAP gates with no quantum cost. In this context, the permutation operation $(0\ 7\ 5\ 14\ 9\ 10\ 15\ 12\ 3\ 8\ 4\ 11) (1\ 13\ 6\ 2)$ can be directly implemented in a 16-qubit quantum circuit (See Fig. ~\ref{fig9}-a). The inverse of the P-box, referred to as the P$^{-1}$-box, undoes the permutations performed by the P-box. As shown in Fig.~\ref{fig9}-b, it is implemented in a similar manner using SWAP gates with zero quantum cost. The permutation operation for the inverse mapping is 
(0 11 4 8 3 12 15 10 9 14 5 7)(1 2 6 13), which corresponds directly to the reverse mapping of the original P-box operation.

\begin{figure}[H]
\centering
\includegraphics[width=0.9\textwidth]{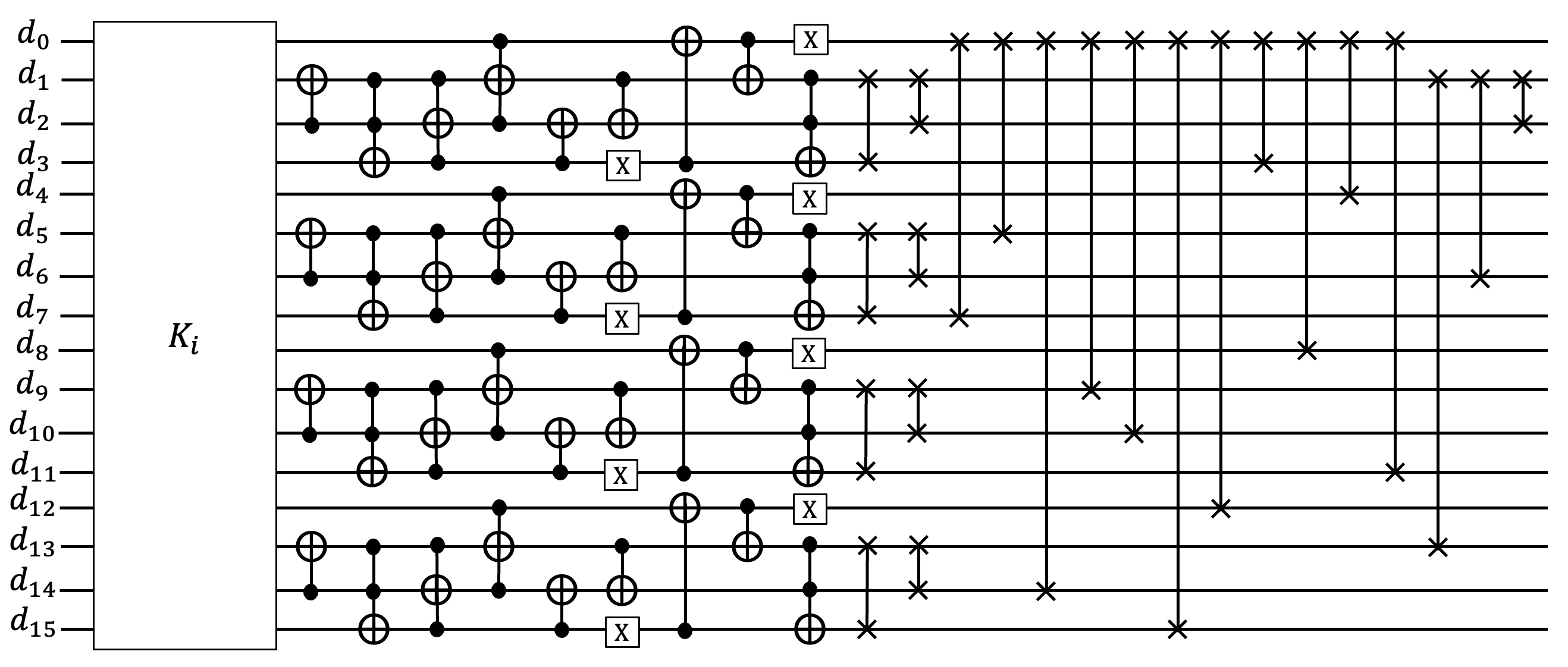}
\caption{\centering Illustration of one round of implementation of KSP} 
\label{fig10}
\end{figure}

\vspace{1em} 
\noindent \textbf{KSP:} We combine the key addition, substitution, and permutation layers into a single cohesive unit for constructing the KSP structure, which plays a critical role in the implementation of the SLIM cipher. For one encryption round, the KSP acts on the 16 qubits coming from the D-box ($D_i$) is given in Fig.~\ref{fig10}.

\noindent First, the 16 qubits from $D_i$ are CNOTed with the round key $K_i$, derived from the key schedule (Fig. ~\ref{fig6}). Following this, the transformed qubits are divided into four 4-qubit blocks, each of which is processed in parallel through the quantum S-box circuits. These S-boxes, optimized for quantum implementation, introduce the required non-linear transformations (Fig. ~\ref{Fig7}). The outputs of the four parallel S-box operations are then routed through the P-layer, which performs a predefined permutation of the qubits (Fig. ~\ref{fig9}). This layer is implemented using quantum SWAP gates, ensuring zero quantum cost for the permutation operation. Upon completing the P-layer, the KSP structure for the current round is finalized, preparing the qubits for interaction with the U-box in the subsequent step.

\noindent This modular design of the KSP not only ensures efficiency but also allows for straightforward reversibility, leveraging the Feistel structure of SLIM. A similar process can be constructed for the inverse operation, KSP$^{-1}$, which includes the inverse permutation layer (P$^{-1}$), the inverse substitution layer (S$^{-1}$), and the reverse key addition layer (K$^{-1}=$K). These components work together to efficiently return the qubits to their initial state, ensuring seamless integration into the overall quantum implementation of SLIM.

\section{Quantum resources for SLIM}\label{sec4}

The total quantum cost of implementing the SLIM algorithm is evaluated by calculating the number of quantum gates required to construct its circuits. Specifically, the cost is determined using standard gate metrics: a NOT gate has a cost of 1 unit, a CNOT gate also costs 1 unit, and a CCNOT gate (Toffoli gate) has a cost of 5 units \cite{shende2009cnot}. The SLIM algorithm operates over 32 rounds in total. As described in Section \ref{sec3}, the first 5 rounds involve the direct use of 80 qubits in the K layer without requiring additional quantum gate operations for key scheduling. Consequently, these rounds incur no extra gate cost. However, starting from the 6th round, the K layer introduces a significant computational overhead due to the quantum gates required for the key scheduling process. Therefore, to calculate the total quantum resource requirements for SLIM, the first 5 rounds and the subsequent rounds should be considered separately (See Table~\ref{tab2}).

According to the quantum algorithm of SLIM (Fig.~\ref{fig6}), each round consists of three main steps: the first half of the qubits undergo KSP, followed by CNOT operations with the second half of the qubits, and then the qubits pass through KSP$^{-1}$. For simplicity, the cost of the CNOT operations between KSP and KSP$^{-1}$ is included in the cost of KSP. The total cost in the first 5 rounds is 1300, which is sum of the costs of KSP and KSP$^{-1}$ as given at 4th and 5th rows in Table \ref{tab2}. In the subsequent rounds, each K step includes 20 NOT, 130 CNOT, and 40 CCNOT. However, the cost of KSP and KSP$^{-1}$ for the remaining 27 rounds is 25920. As a result, the total cost required for the quantum implementation of SLIM is 27220.
 
\begin{table}[h]
\centering
\caption{Quantum resources requirement for the proposed quantum implementation of SLIM}\label{tab2}

\begin{tabular*}{\textwidth}{@{\extracolsep\fill}p{1cm}p{0.7cm}p{0.7cm}p{0.7cm}p{0.7cm}p{0.7cm}p{0.7cm}p{0.7cm}p{0.7cm}p{0.7cm}p{0.7cm}}
\toprule%
LAYER & NOT & CNOT & CCNOT & TOTAL & COST \\
\midrule
S  & 2 & 5 & 4 & 11 & 27 \\
S$^{-1}$  & 4 & 2 & 4 & 10 & 26 \\
$K_{i>5}$  & 20 & 130 & 40 & 5130 & 9450 \\
$KSP_{i\leq5}$  & 8 & 52 & 16 & 380 & 700 \\
$KSP_{i\leq5}^{-1}$  & 16 & 24 & 16 & 280 & 600 \\
$KSP_{i>5}$  & 28 & 182 & 56 & 7182 & 13230 \\
$KSP_{i>5}^{-1}$ & 36 & 154 & 56 & 6642 & 12690 \\
SLIM  & 1848 & 9452 & 3184 & 14484 & 27220 \\
\botrule
\end{tabular*}
\end{table}

\noindent Besides, the depth of a quantum circuit (simply quantum-depth), measures to the number of sequential steps required to execute all operations in the circuit. Unlike classical depth, which refers to the maximum number of gates along a single left-to-right path through the circuit, quantum depth accounts for the layering of quantum gates, where parallelizable gates within a layer are considered as one step. This distinction highlights the unique structure of quantum circuits and their reliance on parallelism to optimize execution. Moreover, certain gates, such as the Toffoli gate, play a critical role in determining quantum depth. While a single Toffoli gate has a T-depth of 1, its full decomposition requires seven layers and four additional ancilla qubits \cite{selinger2013quantum}. This metric is crucial as it reflects both the circuit's computational complexity and the time required for execution. Deeper circuits often indicate more intricate computations but may also introduce challenges in maintaining coherence.

In the quantum implementation of SLIM, the depth of the circuit is determined by the different components used in each round. The depth of the S-box is 33, while its inverse, the S$^{-1}$box, has a depth of 32. For the K layer in rounds $i>5$, the depth is 35. For the first 5 rounds ($i\leq 5$), the combined depth of 
KSP and KSP$^{-1}$ is relatively small, totaling 340. However, in the subsequent 27 rounds ($i>5$), the depth increases as KSP and KSP$^{-1}$ together contribute a total depth of 3,726. Adding these values together, the total depth of SLIM across all 32 rounds is 4,066. On the other hand, the absence of ancilla qubits notably increases the quantum cost. If ancilla qubits were utilized, the need for KSP$^{-1}$ operations would be eliminated. In this scenario, the total cost would comprise 13,930 units from the KSP operations, along with an additional 496 units attributed to the CNOT operations required for managing the ancilla qubits. This adjustment would bring the total quantum cost down to 14,426 units.

A detailed analysis of SLIM, particularly in comparison to other block ciphers, requires an evaluation of its total quantum cost, circuit depth, and overall qubit utilization (see Table~\ref{tab3}). Within the domain of quantum computing, where efficiency and resource optimization are paramount, SLIM demonstrates a well-balanced approach, achieving notable depth efficiency while maintaining competitive performance and resource demands.

\begin{table}[h]
\centering
\caption{Cost and qubit comparison of quantum implementations with other BCs}\label{tab3}
\begin{tabular*}{\textwidth}{@{\extracolsep\fill}p{1.8cm}p{0.8cm}p{0.5cm}p{0.5cm}p{0.6cm}p{0.8cm}p{0.8cm}p{0.8cm}p{0.8cm}p{0.8cm}p{0.5cm}}
\toprule%
CIPHER & QUBIT & BLK & KEY & NOT & CNOT & CCNOT & TOTAL & DEPTH & COST & REF \\
\midrule
SIMON  & 192 & 64 & 128 & 1216 & 7396 & 1408 & 10020 & 2643 & 15652 & \cite{anand2020} \\
SIMON  & 256 & 128  & 128  & 4224 & 17152 & 4352  & 25728 & 8427 & 43136 & \cite{anand2020}  \\
RECTANGLE  & 144 & 64  & 80  & 567 & 4964 & 2000  & 7531 & 226 & 15531 & \cite{saravanan2021} \\
LBLOCK  & 144 & 64  & 80  & 877 & 16747 & 14280  & 31904 & 1740 & 89024 & \cite{jing2023} \\
LiCi  & 192 & 64  & 128  & 379 & 12900 & 8624  & 21903 & 1210 & 56399 & \cite{jing2023} \\
PUFFIN  & 192 & 64  & 128  & 620 & 3136 & 3584  & 7340 & 353 & 21676 & \cite{paramasivam2023} \\
PRINT  & 128 & 48  & 80  & 154 & 3840 & 2304  & 6298 & 336 & 15514 & \cite{paramasivam2023} \\
SM4  & 260 & 32  & 128  & 8750 & 134912 & 26624  & 170286 & 20480 & 276782 & \cite{luo2024quantum} \\
SLIM  & 112 & 32  & 80  & 1848 & 9452 & 3184  & 14484 & 4066 & 27220 & This work \\
\botrule
\end{tabular*}
\end{table}

\section{Conclusion}\label{sec13}

In this study, a quantum implementation for SLIM has been proposed for the first time, offering a lightweight BC design optimized for quantum cost. SLIM distinguishes itself as the algorithm with the lowest qubit requirement (112 qubits) compared to other BCs in the literature (see Table~\ref{tab3}). This compact qubit utilization, combined with a well-balanced quantum cost of 27,220, positions SLIM as a resource-efficient solution in quantum cryptography. Its unique design eliminates the need for ancilla qubits, simplifying its implementation while maintaining its robustness.

The quantum implementation of SLIM aligns with the broader goal of developing encryption algorithms that are resilient to future quantum threats. By thoroughly analyzing the S, P, and K layers, this study provides the first comprehensive evaluation of SLIM's behavior in the quantum domain. The results reveal a total gate depth of 4,066, highlighting its efficiency compared to other block ciphers, as many alternatives require significantly higher depths for similar levels of security. SLIM's efficient gate distribution is evidenced in Table~\ref{tab3}, where critical layers such as KSP and KSP$^{-1}$ are shown to manage their computational demands effectively without relying on additional quantum resources.

What sets SLIM apart is its balance between resource requirements and performance. While other quantum implementations, such as SIMON, RECTANGLE, and LBLOCK, achieve varying degrees of success, they require higher qubit counts or impose larger quantum costs (see Table~\ref{tab2}). In contrast, SLIM minimizes its quantum footprint while maintaining competitive cryptographic security. Its low qubit count, optimized quantum cost, and acceptable depth make it a strong candidate for lightweight encryption in the era of quantum computing.

In conclusion, this research highlights SLIM's strengths as a quantum-resistant lightweight block cipher. It marks an important step toward addressing the vulnerabilities of classical cryptographic methods in the face of advancing quantum computing. By achieving a notable balance between qubit efficiency, quantum cost, and circuit depth, SLIM provides a promising framework for secure encryption protocols in a quantum-enabled future. Continued research will further enhance its resilience, paving the way for robust cryptographic solutions to secure sensitive data against the dynamic landscape of quantum threats.

\section*{Data Availability}

All data generated or analysed during this study are included in this published article and its supplementary information files.

\backmatter

\bmhead{Funding} Not applicable
\bmhead{Conflict of interest/Competing interests}
\bmhead{Ethics approval and consent to participate}
\bmhead{Consent for publication}

\bibliography{sn-bibliography}

\end{document}